\documentclass{osa-article}

\journal{oe}
\articletype{Research Article}
\begin{document}

\title{Hybrid III-V diamond photonic platform for quantum nodes based on neutral silicon vacancy centers in diamond}

\author{Ding Huang,\authormark{1,\dag} Alex Abulnaga,\authormark{1,\dag} Sacha Welinski,\authormark{1} Mouktik Raha,\authormark{1} Jeff D. Thompson,\authormark{1} and Nathalie P. de Leon\authormark{1,*}}

\address{\authormark{1}Department of Electrical Engineering, Princeton University, New Jersey 08544, USA}
\address{\authormark{\dag}These authors contributed equally to this work.}

\email{\authormark{*}npdeleon@princeton.edu} %% email address is required



\begin{abstract}
Integrating atomic quantum memories based on color centers in diamond with on-chip photonic devices would enable entanglement distribution over long distances. However, efforts towards integration have been challenging because color centers can be highly sensitive to their environment, and their properties degrade in nanofabricated structures. Here, we describe a heterogeneously integrated, on-chip, III-V diamond platform designed for neutral silicon vacancy (SiV$^0$) centers in diamond that circumvents the need for etching the diamond substrate. Through evanescent coupling to SiV$^0$ centers near the surface of diamond, the platform will enable Purcell enhancement of SiV$^0$ emission and efficient frequency conversion to the telecommunication C-band. The proposed structures can be realized with readily available fabrication techniques. 
\end{abstract}

%%%%%%%%%%%%%%%%%%%%%%%%%%  body  %%%%%%%%%%%%%%%%%%%%%%%%%%
\section{Introduction}

Color centers in diamond combine optical addressability with long spin coherence times, making them promising candidates for repeater-based quantum networks ~\cite{atature2018material}. Recent proof-of-principle demonstrations of key ingredients for quantum networks using color centers in diamond include on-demand remote entanglement generation ~\cite{humphreys2018deterministic}, coherent control of multiple nearby nuclear spin memories ~\cite{bradley2019ten}, entanglement distillation ~\cite{kalb2017entanglement}, and memory-enhanced quantum communication ~\cite{bhaskar2020experimental}. However, building larger quantum networks that incorporate more than two nodes and span longer distances will require new technological breakthroughs. Currently known platforms suffer from either degraded optical coherence when incorporating into nanophotonic devices ~\cite{chu2014coherent, ruf2019optically}, or limited electron spin coherence times of milliseconds combined with the requirement to work at millikelvin temperatures ~\cite{sukachev2017silicon}. The recently reported SiV$^0$ center in diamond has the potential to overcome many of these challenges ~\cite{rose2018observation,green2017neutral,rose2018strongly,zhang2020optically}. The unique combination of stable optical transitions and long spin coherence times at liquid helium temperature makes the SiV$^0$ center an attractive building block for nodes in quantum networks. 

One proposal for enhancing the entanglement generation rate in color-center-based quantum networks is to integrate color centers with nanophotonic devices. In particular, optical cavities greatly enhance atom-photon interaction, which improves spin readout and spin-photon entanglement fidelity. Recent progress in diamond nanofabrication techniques has enabled record single-emitter cooperativities up to 105 ~\cite{bhaskar2020experimental}, as well as interfaces with multiple color centers ~\cite{evans2018photon}. Furthermore, nanophotonic devices can enable other functionality such as on-chip quantum frequency conversion (QFC), which is key to achieving long-distance quantum communication. The small mode volume offers an opportunity for highly efficient nonlinear optical interactions even with low pump powers, which increases the signal-to-noise ratio for single photon level signals \cite{li2016efficient}.

Monolithic fabrication techniques of diamond nanophotonic cavities require milling or etching bulk single crystal diamond, which leads to low yield ~\cite{nguyen2019integrated} and precludes the on-chip integration of photonic components with other functionality, such as QFC and active devices. Furthermore, there is currently no method for high purity, wafer-scale synthesis of single crystal diamond ~\cite{nelz2019toward}, limiting the scalability of this approach.

By contrast, nanofabrication techniques in III-V semiconductor systems are fairly mature. Nanophotonic components with diverse functionalities have been fabricated in III-V semiconductors such as gallium arsenide (GaAs) \cite{dietrich2016gaas} and gallium phosphide (GaP) \cite{Wilson2020}. GaAs photonic crystal cavities with quality factors exceeding $10^4$ have been demonstrated at wavelengths close to the zero-phonon line of the SiV$^0$ center ($\sim$ 946 nm) \cite{hennessy2007quantum, englund2007controlling}. More recently, surface passivation schemes helped to push the quality factors of state-of-the-art cavities to be above $10^5$ \cite{kuruma2020surface}. Other recent efforts in GaP photonics have been spurred by its large optical nonlinearities and transparency over a wide wavelength range ~\cite{Wilson2020}. GaP cavities with high quality factors have been reported using microdisks ~\cite{mitchell2014cavity}, one-dimensional photonic crystal cavities ~\cite{schneider2019optomechanics} and ring resonators ~\cite{Wilson2020}. 

\begin{figure}[ht!]
\centering\includegraphics[width=\textwidth]{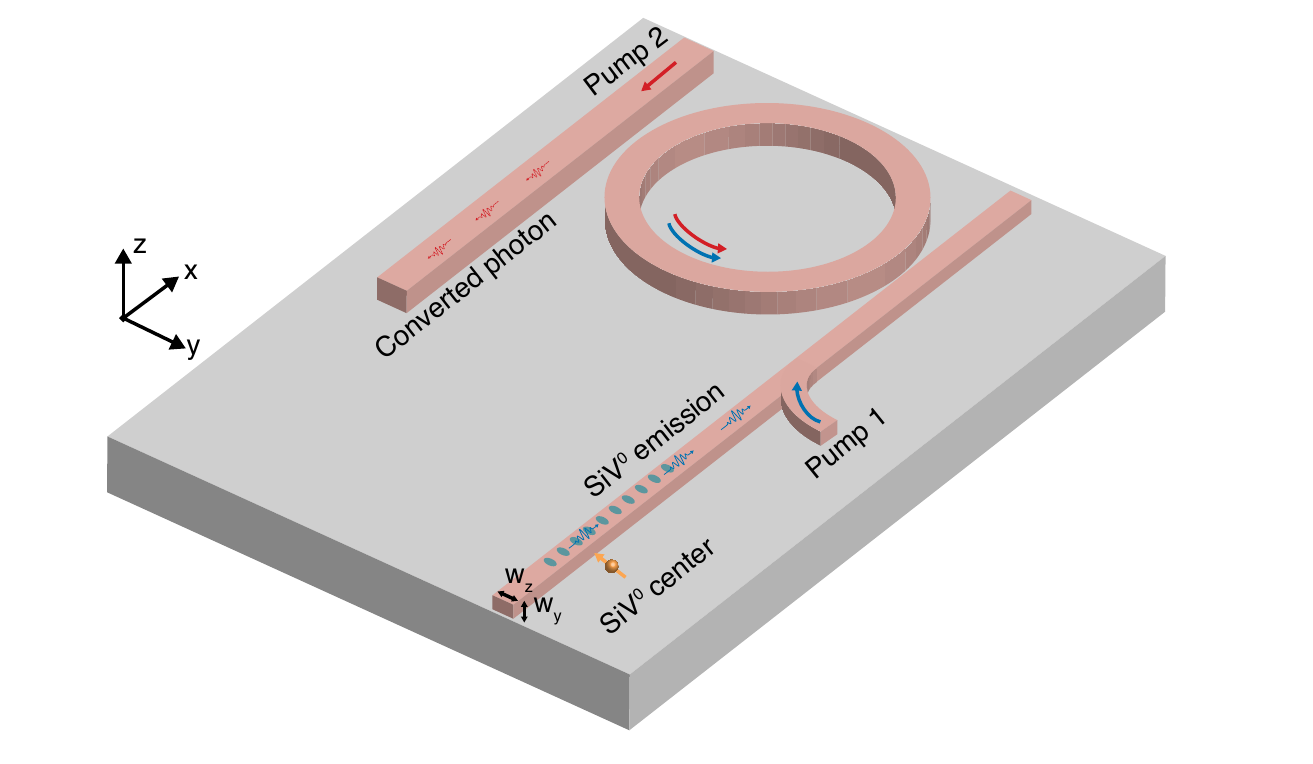}
\caption{Schematic illustration of the hybrid III-V diamond photonic platform. III-V material (pink) is patterned on top of the diamond substrate (grey). The SiV$^0$ center near the diamond surface is evanescently coupled to a nanobeam photonic crystal cavity with a width of $w_\text{y}$ and a thickness of $w_\text{z}$. The emitted photon is routed to an on-chip frequency conversion module, where FWM-BS scheme is used to translate the signal to single photon at the telecommunication C-band.}
\label{fig:schematic}
\end{figure}

A promising method to mitigate the constraints imposed by diamond nanofabrication is heterogeneous integration of diamond and a separate device layer material ~\cite{wan2019large}. Instead of fabricating devices directly on single crystal diamond, the photonic device is fabricated in a high-index photonic layer on top of the diamond substrate such that photons can evanescently couple to color centers that are close to the diamond surface (Fig.\ref{fig:schematic}). This scheme has previously been used to demonstrate Purcell enhancement of optical emission from the negatively charged nitrogen vacancy (NV) center in diamond ~\cite{englund2010deterministic, barclay2009hybrid,gould2016efficient} and Er$^{3+}$ ions in yttrium orthosilicate ~\cite{dibos2018atomic, raha2020optical, chen2020parallel}. For NV centers, previous work has involved fabricating GaP waveguides or microdisks on top of diamond, and then subsequently etching into the diamond to realize a high quality factor structure ~\cite{barclay2009hybrid, barclay2011hybrid, gould2016efficient}. This subsequent etching step leads to significant spectral diffusion of the NV center optical transition \cite{ruf2019optically, schmidgall2018frequency}.

Here, we develop a design for heterogeneously integrated nanophotonic devices to build quantum nodes based on the SiV$^0$ center, using a hybrid III-V diamond platform. One-dimensional photonic crystal cavities enhance the optical emission from single SiV$^0$ centers, and the emission can then be routed on-chip to a microresonator-based frequency converter based on a four-wave mixing Bragg scattering (FWM-BS) scheme. In contrast to previous demonstrations, our design does not require etching into the diamond, avoiding deleterious effects on the color center. 

\section{Photonic crystal cavity design}

Photonic crystal cavities provide mode confinement to a small volume on the order of a cubic wavelength, enabling enhanced coupling between a single color center and the cavity mode \cite{lodahl2015interfacing}. The key figure of merit is the Purcell enhancement factor of the spontaneous emission rate, $P$, which is given by $P = 4g(\vec{r})^2/\kappa \Gamma_0$. Here, the rate of interaction between the cavity mode and a single SiV$^0$ center at the position $\vec{r}$ is characterized by the single-photon Rabi frequency, $g(\vec{r)} = \vec{\mu}\cdot\vec{E}(\vec{r})/\hbar$, where $\vec{\mu}$ is the electric dipole moment of the optical transition for the SiV$^0$ center and $\vec{E}(\vec{r})$ is the electric field strength at the position of the SiV$^0$ center associated with a single photon in the cavity mode. The cavity decay rate, $\kappa$, is related to the quality factor of the cavity, Q, and the cavity resonance frequency, $\omega$, by $\kappa = \omega/Q$. The spontaneous emission rate of the optical transition of interest, $\Gamma_0$, is an intrinsic property of the SiV$^0$ center that can be characterized independent of the cavity.

\begin{figure}[ht!]
\centering\includegraphics[width=\textwidth]{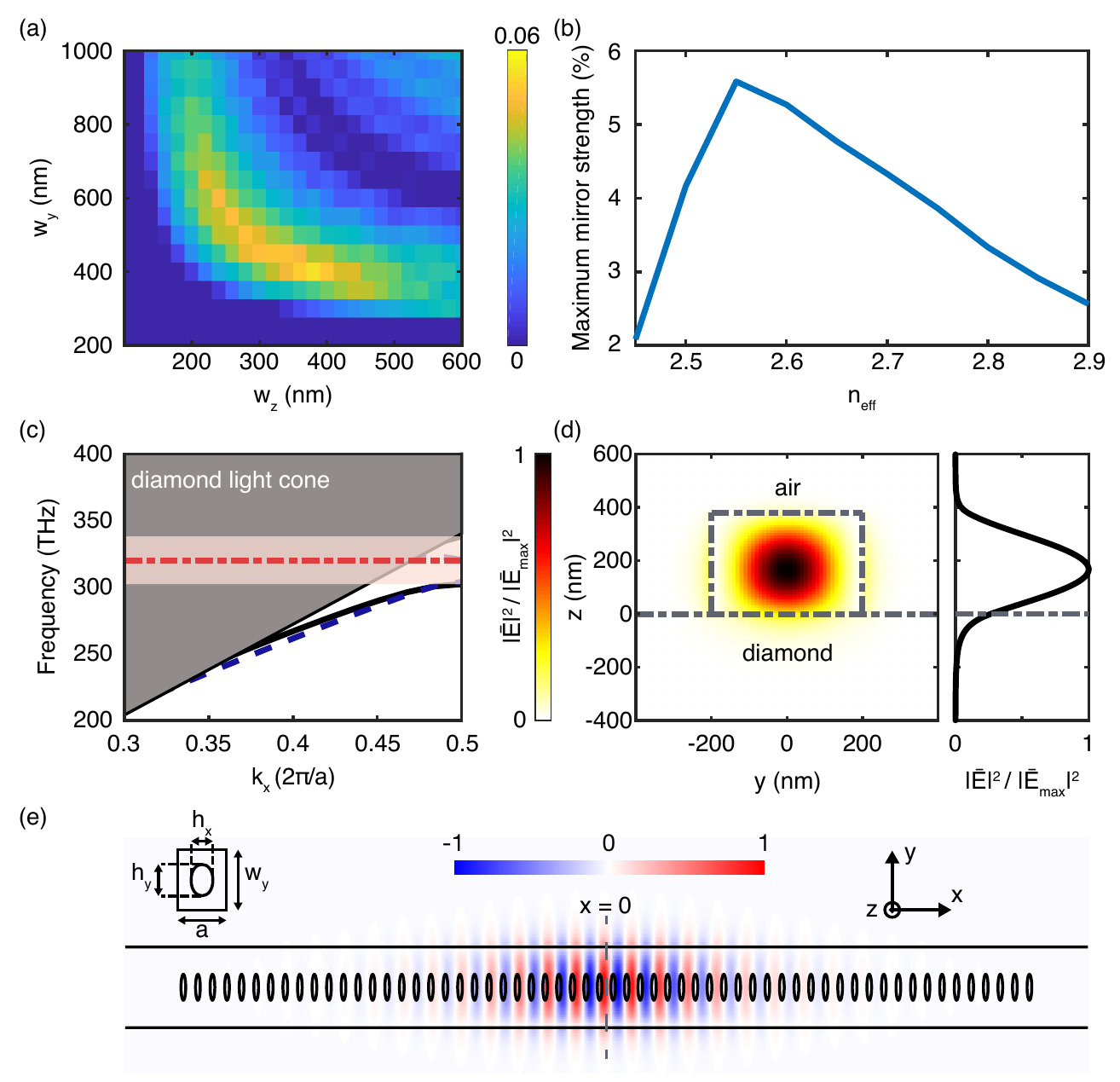}
\caption{Design of a GaP nanobeam cavity on a diamond substrate. (a) Mirror strength for different values of $w_\text{y}$ and $w_\text{z}$ with $n_{\text{eff}} = 2.55$. (b) Maximum mirror strength at different $n_{\text{eff}}$ values. (c) Band diagram for a nanobeam waveguide with $(w_\text{y}, w_\text{z}, h_\text{x}, h_\text{y}, a) = (400, 380, 59, 128, 184)$ nm. TE modes and TM modes are indicated by solid black lines and dashed blue lines respectively. The region above the diamond light line is shaded in grey and the fundamental TE bandgap is indicated by the pink shaded area. The target frequency is marked by the dashed red line. (d) Normalized electric field $|E|^2$ distribution (left) in the yz-plane at x = 0 with a line cut of the electric field at y = 0 nm (right). (e) Normalized $E_\text{y}$ ﬁeld proﬁle of the fundamental cavity mode of the GaP nanobeam cavity. The xy slice is taken at the center of the cavity (i.e. z = 190 nm).}
\label{fig:mpb_design}
\end{figure}

This points to two parameters to optimize to build an efficient spin-photon interface: high Q and concentrated electric field with a good overlap between the color center and the cavity mode to maximize $\vec{E}(\vec{r})$. Here, we optimize the design for a heterogeneously integrated GaP-on-diamond photonic crystal cavity. Suspended GaP photonic crystal cavities with high Q resonances at telecommunication wavelengths have been demonstrated ~\cite{schneider2019optomechanics}. The large bandgap of GaP minimizes absorption losses at near-infrared wavelengths and allows for cavities with high Q. The diamond substrate has a moderately high refractive index ($n_\text{Dia} \sim 2.4$), presenting a design challenge for achieving high Q. In contrast to suspended photonic crystal devices, a diamond substrate expands the light cone into which photons can radiate out of the cavity [Fig.\ref{fig:mpb_design}(c)]. We propose a one-dimensional photonic crystal nanobeam cavity, where high Q resonances can be supported. For realistic beam width and height, the proposed heterogeneously integrated structure supports resonant modes with quality factors $\sim 10^{5}$, on par with state-of-the-art freestanding GaP photonic crystal cavities.

We first consider a GaP nanobeam with waveguide width $w_\text{y}$ and waveguide thickness $w_\text{z}$. As illustrated by Fig.\ref{fig:schematic}, the high refractive index of GaP ($n_\text{GaP} \sim 3.13$) provides the optical mode confinement along the y- and z-direction. The cavity is defined by a one-dimensional lattice of elliptical air holes with lattice constant $a$ and diameters $h_\text{x}$ and $h_\text{y}$ along the x- and y-directions respectively [Fig.\ref{fig:mpb_design}(e), inset]. Starting with the design that maximizes the bandgap, the nanobeam cavity is formed by locally perturbing the periodic array to form a defect \cite{quan2011deterministic}. An adiabatic shift of the band-edge frequencies moves the target frequency from the dielectric band-edge at the defect to the middle of bandgap in the Bragg mirror regions. This can be achieved by either tapering the size of the holes or the lattice constant. As an example, we employ the latter approach to design a symmetric cavity where the lattice constant is gradually increased from $a_\text{cav}$ in the middle of the cavity to $a_\text{mirr}$ in the mirror regions on both sides.

To find the design with maximum mirror strength, we choose a lattice constant according to $a = \lambda_0/2n_{\text{eff}}$, where $\lambda_0$ is the target wavelength and $n_{\text{eff}}$ is the effective mode index of the cavity. In the case of the GaP-on-diamond system, the effective indices of the cavity modes have to lie in between the refractive index of diamond and that of GaP, i.e. $n_{\text{eff}} \in (2.4,3.13)$. Using MIT Photonic Bands ~\cite{johnson2001block}, we calculate the band structure of the GaP nanobeam on a diamond substrate for different structural parameters. As modes above the diamond light line are unconfined, we are only interested in those below the light line at $k_\text{x} = \pi / a$. 

We define the mirror strength as the separation between the target frequency and the nearest quasi-transverse-electric (TE) guided mode normalized by the target frequency. The TE modes and the quasi-transverse-magnetic (TM) modes are defined as modes with most of their electric field components along the y-axis and the z-axis respectively. Fig.\ref{fig:mpb_design}(a) shows the largest mirror strength at different combinations of $w_\text{y}$ and $w_\text{z}$ for $n_{\text{eff}} = 2.55$. At each point on the two-dimensional sweep, we optimize the mirror strength with different combinations of hole diameters (i.e. $h_\text{x}$ and $h_\text{y}$). By repeating the two-dimensional sweep for different $n_{\text{eff}}$ values, we could identify the cavity design with the maximum mirror strength at the target frequency. Fig.\ref{fig:mpb_design}(b) shows the largest mirror strength for different $n_{\text{eff}}$ values. As $n_{\text{eff}}$ gets closer to the refractive index of diamond ($n_\text{Dia} \sim 2.4$), the mirror strength is limited by leaky modes above the diamond light line. On the other hand as $n_{\text{eff}}$ increases, more modes are pushed beneath the light line, and the mirror strength decreases as the modes are pushed closer together. The optimal design (i.e. $n_{\text{eff}} = 2.55$) has an air band-edge frequency slightly below the diamond light line with structural parameters $(w_\text{y}, w_\text{z}, h_\text{x}, h_\text{y}, a)$ = (400, 380, 59, 128, 184) nm. For the cavity design with maximum mirror strength, we plot the band diagram for different modes in Fig.\ref{fig:mpb_design}(c). The desired resonant frequency is in the middle of the fundamental TE bandgap. We achieve a bandgap of $\sim 11\%$.

In our cavity design, we keep structural parameters $w_\text{y}$, $w_\text{z}$, $h_\text{x}$ and $h_\text{y}$ to be the same across the entire nanobeam. By varying the lattice constant from $a_\text{cav}$ = 171 nm in the middle of the cavity to $a_\text{mirr}$ = 184 nm in the mirror region, we adiabatically move the dielectric band-edge such that the target resonance frequency is in the middle of the fundamental TE bandgap.

The Purcell enhancement factor depends on the overlap between the transition dipole moment for the SiV$^0$ center and the resonant mode. To examine the efficacy of our cavity design approach, we study the cavity mode profile using three-dimensional finite-difference time-domain (FDTD) simulations. As an example, we perform FDTD simulations on a cavity where the lattice constant is parabolically tapered from $a_\text{mirr}$ to $a_\text{cav}$ across 10 holes on a single side (20 holes in total in the cavity region). In order to further increase the mirror strength, we add 20 Bragg mirror holes on both ends of the cavity. The normalized $E_\text{y}$ ﬁeld components for the fundamental localized cavity mode are plotted in Fig.\ref{fig:mpb_design}(e). Most of the field strength is concentrated in the adiabatically tapered cavity region. The electric field strength at the position of the SiV$^0$ center varies depending on its location in the diamond substrate [Fig.\ref{fig:mpb_design}(d), left]. At around 50 nm into the diamond substrate, where we observed SiV$^0$ centers with stable optical properties and long spin coherence \cite{rose2018observation}, the electric field energy density, $|E(\vec{r})|^2$, is $9.6\%$ of its maximum value in the center of the cavity [Fig.\ref{fig:mpb_design}(d), right].

\begin{figure}[ht!]
\centering\includegraphics[width=\textwidth]{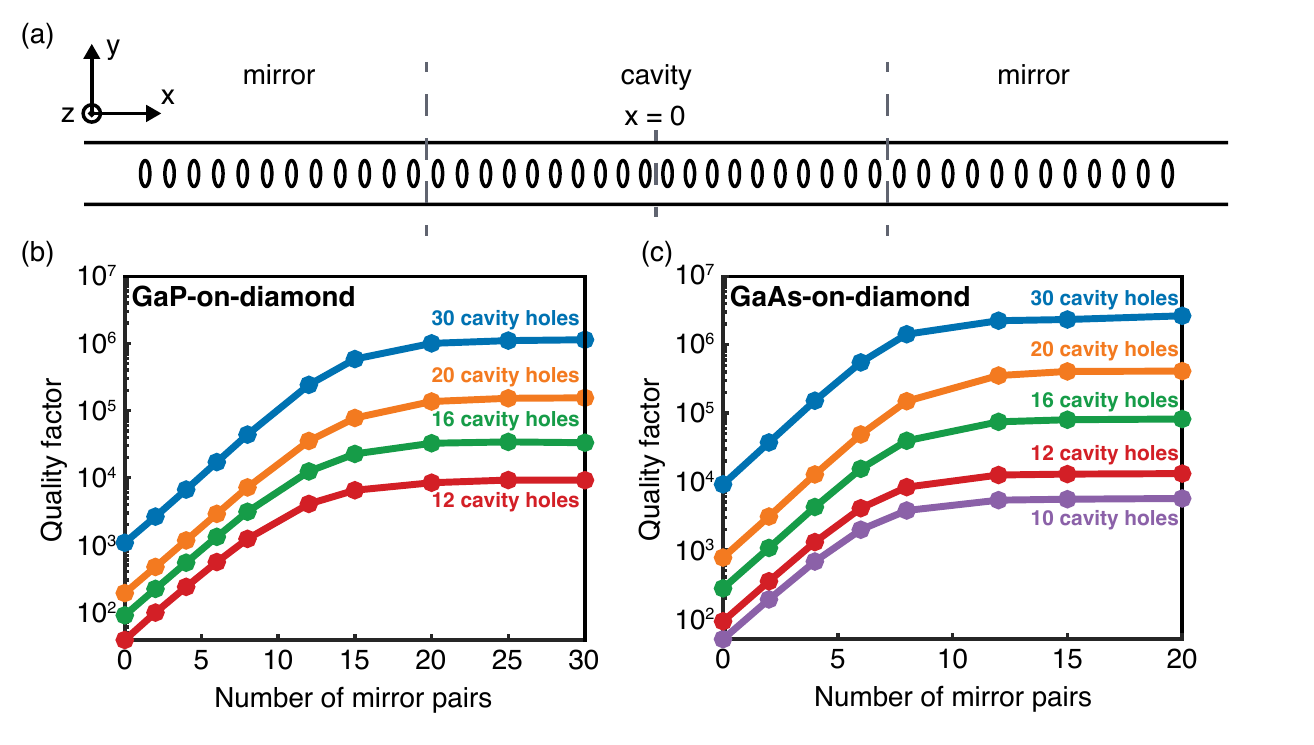}
\caption{(a) Geometry of the symmetric photonic crystal cavity design. In the cavity region, the hole spacing is adiabatically tapered from $a_\text{cav}$ to $a_\text{mirr}$. In the mirror regions on both sides, the hole spacing is held constant at $a_\text{mirr}$. The schematic shows a design with 20 cavity holes and 12 mirror pairs. Dependence of the GaP nanobeam (b) and the GaAs nanobeam (c) cavity Q on the number of mirror pairs for 10 (purple), 12 (red), 16 (green), 20 (orange) and 30 (blue) holes in the cavity region.}
\label{fig:FDTD_sim}
\end{figure}

We evaluate the scaling of Q with different design parameters using 3D FDTD. The nanobeam cavity can be divided into a cavity region and two Bragg mirror regions [Fig.\ref{fig:FDTD_sim}(a)]. In the cavity region, the lattice constant tapers parabolically from $a_\text{cav}$ in the middle to $a_\text{mirr}$ on both ends. As the cavity is symmetric with respect to x = 0, the total number of holes in the entire cavity region is twice the number of holes on a single side. For simplicity, we will refer to different designs based on the total number of holes in the cavity region in the following discussions. The mirror segments consist of periodic arrays of holes where the lattice constant is fixed at $a_\text{mirr}$. Since our cavity design is symmetric, an additional mirror hole on one end of the cavity implies an extra mirror hole on the opposite side. We define the addition of a pair of Bragg mirror holes as an additional "mirror pair".

First we calculate the Q for different cavity designs. Fig.\ref{fig:FDTD_sim}(b) shows the scaling of Q with the number of mirror pairs. For small numbers of holes in the mirror regions, the Q of the cavity increases with the mirror strength. At higher numbers of mirror holes, Q is limited by radiation loss in the cavity region. In order to further push up the quality factor, we taper the lattice constant over more holes in the cavity region. For 30 cavity holes, the Q is saturated over $10^6$ with 20 mirror pairs, and for 20 cavity holes, the saturated Q exceeds $10^5$, which is the current demonstrated state of the art in suspended GaP cavities ~\cite{schneider2019optomechanics}.

Our method for designing hybrid photonic platforms can be applied to other material systems with a higher refractive index than diamond. Using the aforementioned design principles, we demonstrate that a similar evanescent coupling scheme can be realized using GaAs-on-diamond. Epitaxially grown GaAs has a higher refractive index ($n_\text{GaAs} \sim 3.55$) compared to GaP. The optimized GaAs nanobeam cavity has structural parameters $(w_\text{y}, w_\text{z}, h_\text{x}, h_\text{y}, a_\text{cav}, a_\text{mirr})$ = (350, 220, 70, 136, 162, 180) nm with a bandgap of $\sim 15.7\%$ in the mirror region. Using 3D FDTD simulations, we can similarly evaluate the scaling of Q for different GaAs photonic crystal cavity designs [Fig.\ref{fig:FDTD_sim}(c)]. We find that for the optimized structure, a smaller number of mirror pairs is required to saturate the Q because the mirror strength is larger.

For the optimized cavity designs, we can evaluate the expected Purcell factor for SiV$^0$ located at different locations. After taking local-field correction of the spontaneous emission rate into account \cite{dung2006local}, we can extract the dipole moment for the SiV$^0$ center ($\vec{\mu}$) using the following expression:

\begin{equation} \label{spontaneous_decay}
    \Gamma_0 = \frac{1}{\beta} \cdot \left ( \frac{3n_{\text{Dia}}^2}{2n_{\text{Dia}}^2+1} \right )^2 n_{\text{Dia}}
    \cdot \frac{|\vec{\mu}|^2 \omega^3}{3\pi \epsilon_0 \hbar c^3}
\end{equation}

\noindent where $\beta$ is the fraction of the decay rate caused by spontaneous emission at the transition coupled to the cavity.

\begin{figure}[ht!]
\centering\includegraphics[width=\textwidth]{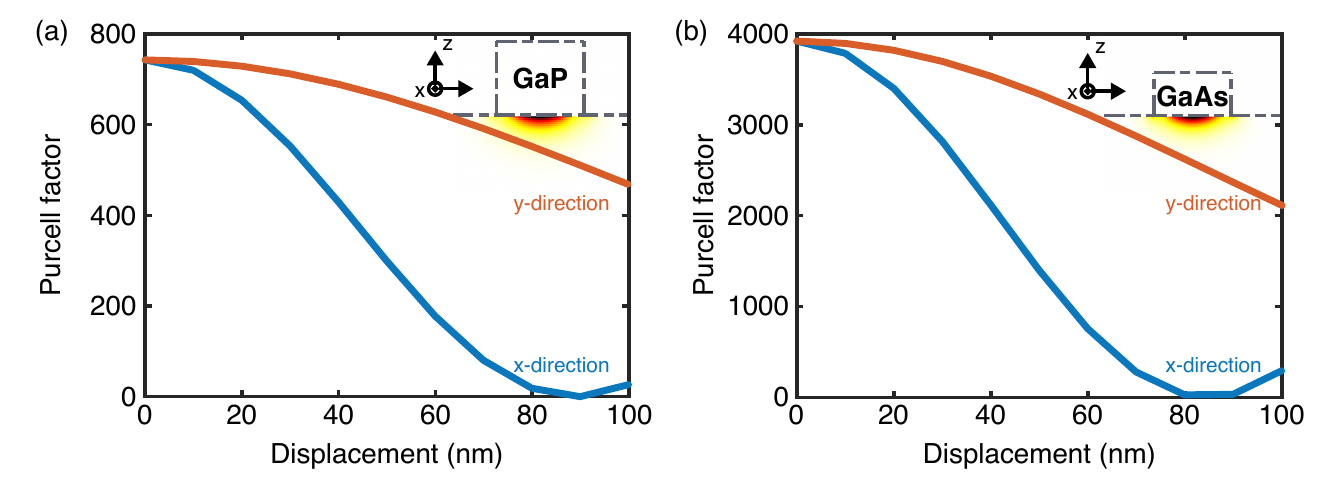}
\caption{Variations of expected Purcell factors for GaP cavity design with 20 cavity holes and 20 mirror pairs (a) and GaAs cavity design with 20 cavity holes and 12 mirror pairs (b). At a depth of 50 nm in the diamond substrate, the maximum attainable Purcell factor for GaP design is 743 and that for GaAs design is 3921. The displacements, in the x-direction (blue) and y-direction (orange), are defined relative to the optimal position. (Inset) Distribution of Purcell factor inside the diamond substrate, y-z plane at x = 0.}
\label{fig:Purcell_variation}
\end{figure}

Here, we make the assumptions that the SiV$^0$ center can be treated as a perfect two-level atom and $\beta = 1$. Using the bulk spontaneous emission rate for SiV$^0$ centers $\Gamma_0 = 2\pi \times 88$ MHz \cite{rose2018observation}, we arrive at the calculated $|\vec{\mu}| = 6.027 \times 10^{-29}$ C-m. If we further assume that the dipole moment is parallel to the electric field polarization of the cavity resonant mode, we can estimate the Purcell factor. The inset of Fig.\ref{fig:Purcell_variation}(a) shows the distribution of expected Purcell factor inside the diamond substrate for GaP-on-diamond. At a depth of 50 nm, the maximum Purcell factor is 743 for a GaP nanobeam cavity with 20 cavity holes and a saturated quality factor of $\sim 10^5$. We also study the sensitivity of Purcell factor to displacements in the xy-plane. The maximum Purcell factor can be achieved when the SiV$^0$ is placed directly under the electric field energy density maximum. As Fig.\ref{fig:Purcell_variation}(a) shows, the Purcell factor is more sensitive to misalignments along the nanobeam direction (i.e. x-direction). Similar results can be obtained for a GaAs nanobeam cavity with 20 cavity holes and a saturated quality factor of $\sim 3.5 \times 10^5$ [Fig.\ref{fig:Purcell_variation}(b)]. As we will discuss in Section \ref{section:discussion}, a Purcell factor above 200 for GaP-on-diamond and a Purcell factor above 1000 for GaAs-on-diamond are practical and attainable using widely available fabrication techniques.

%%%%%%%%%%%%%%%%%%%%%%%%%%%%%%%%%%%%%%%%%%%%%%%%%%%%%%%%%%%%%%%%%
%Frequency Conversion
\section{On-chip quantum frequency conversion} \label{section:QFC}
Low-loss optical fibers provide a scalable way to connect nodes in a long-distance quantum network. To utilize the fiber network, the emission wavelengths of color centers in diamond need to be shifted to overlap with the transparency window of silica fibers. QFC of single photons requires simultaneously achieving high conversion efficiency and low noise. Despite these challenges, recent demonstrations were able to show that the spin-photon entanglement is preserved after the QFC process for NV centers in diamond ~\cite{Dreau2018,Tchebotareva2019}. These experiments used free-space optics in combination with centimeter-long, second-order optically nonlinear ($\chi^{(2)}$) waveguides to perform spectral translations of single photons from the NV emission wavelength (637 nm) to the telecommunication L-band (1588 nm). In addition to leveraging the benefits of low-loss optical fibers, QFC presents a solution to inhomogeneous broadening of color center emission. By detuning the pump laser, the disparate color center emission wavelengths can be shifted to the same telecommunication wavelength, and therefore identical telecom photons can be generated from multiple color-center-based quantum nodes.

\begin{figure}[ht!]
\centering\includegraphics[width=\textwidth]{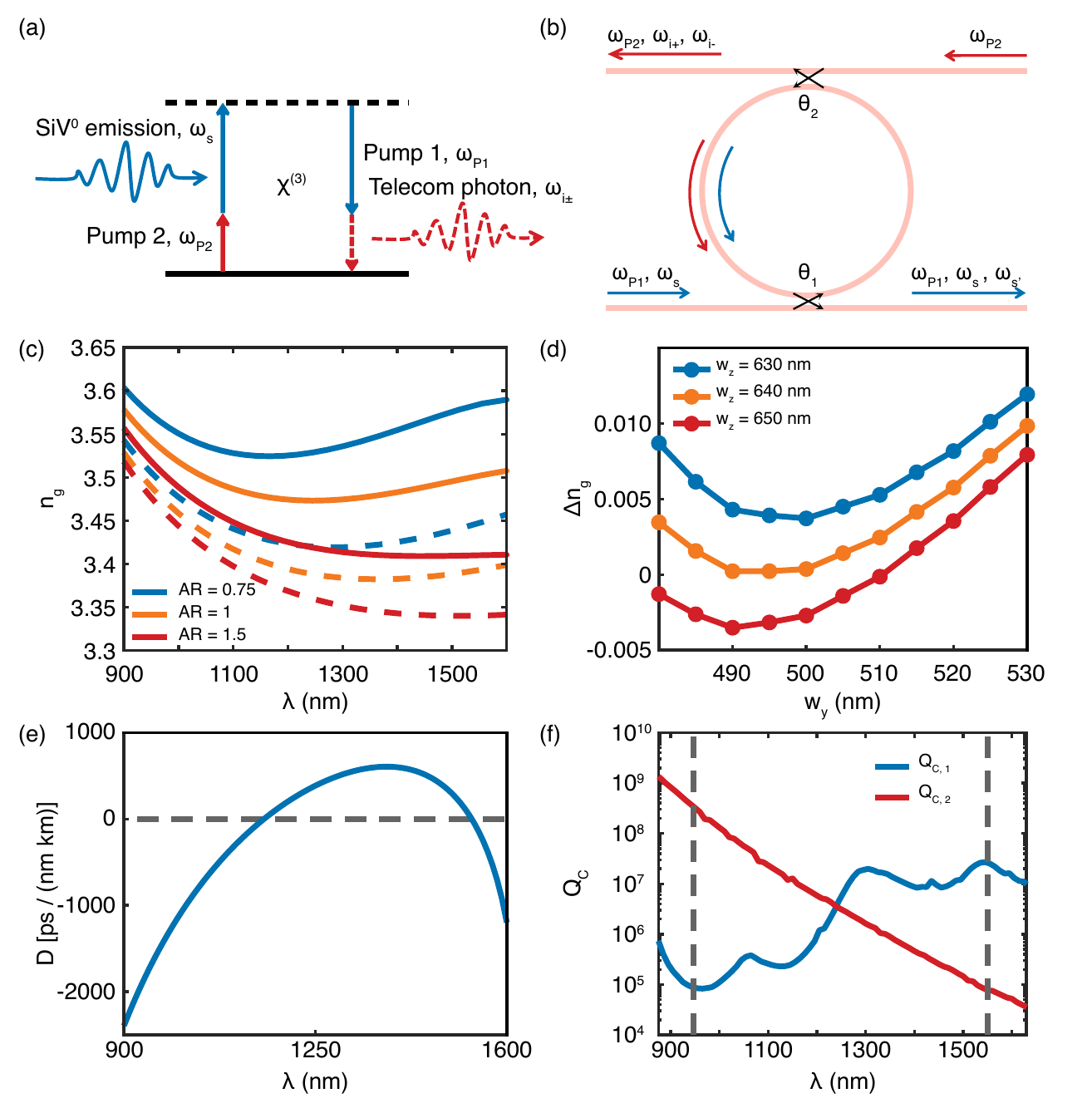}
\caption{(a) Four-wave-mixing Bragg-scattering frequency conversion scheme. The two pump fields, $\omega_\text{p1}$ at 946 nm, and $\omega_\text{p2}$ at 1550nm, combine in a $\chi^{(3)}$ medium to scatter the input SiV$^0$ photon ($\omega_\text{s}$) to idlers $\omega_\text{i$\pm$}$ in the 1550 nm band. (b) The QFC module consists of a microring resonator with two point-couplers to separately couple 946 nm and 1550 nm TE modes. A high-Q resonator allows for prolonged interactions between the  fields in a small mode volume, leading to enhanced nonlinear interactions. (c) Group index of the TE modes as a function of wavelength for different waveguide areas and ARs. Solid lines represent a total cross-sectional area of $A=0.35\mu m^2$, dashed lines are for $A=0.55\mu m^2$. For each cross-sectional area, ARs of 0.75 (blue), 1 (orange) and 1.5 (red) are plotted. (d) Simulated group index mismatch $\Delta n_\text{g} = n_\text{g}^{946} - n_\text{g}^{1550}$ as a function of resonator width, $w_\text{y}$ and thickness, $w_\text{z}$. We achieve $\Delta n_\text{g}=0$ for several cross-sections, and the geometries are relatively insensitive to small variations in width and thickness. (e) Dispersion of the selected ring resonator cross-section ($w_\text{z} = 640$ nm, $w_\text{y}  = 500$ nm) with zero-crossings, i.e. $\text{D}=0$, at 1155 nm and  1536 nm. (f) Coupling quality factor as a function of wavelength for two separate point couplers. The red line represents a 640 nm $\times$ 500nm coupling waveguide at a gap of 210 nm. The blue line represents a 380 nm $\times$ 300 nm coupling waveguide at a gap of 10 nm.}
\label{fig:QFC_intro}
\end{figure}

We propose a compact and power-efficient QFC scheme that can be realized on the hybrid III-V diamond platform. Leveraging the large third order optical nonlinearities ($\chi^{(3)}$) of III-V semiconductors such as GaAs and GaP, we show that a microresonator-based FWM-BS scheme can efficiently translate the emitted photons from SiV$^0$ centers to the telecommunication C-band (1550 nm). In the FWM-BS scheme, two pump fields, $\omega_\text{p1}$ and $\omega_\text{p2}$, create an effective modulation of $\chi^{(3)}$ and scatter an input signal photon ($\omega_\text{s}$) to the idler frequency $\omega_\text{i}$ [Fig.\ref{fig:QFC_intro}(a)]. In principle, this enables noise-free frequency translation of the signal photon by an amount set by the difference between the two pump frequencies \cite{mckinstrie2005translation}. QFC based on FWM-BS has recently been demonstrated to convert the emission of InAs/GaAs quantum dots using Si$_\text{3}$N$_\text{4}$ micro-ring resonators \cite{Singh2018}. Owing to their large $\chi^{(3)}$, III-V semiconductors are promising material platforms in realizing efficient frequency conversion with low pump powers \cite{chang2020ultra, Wilson2020}. Here, we extend the functionalities of the hybrid III-V diamond system and outline our design of an integrated QFC module based on GaP.

Because nonlinear optical effects are generally quite weak, appreciable conversion efficiency requires long interaction times between the pumps and the signal, as well as high pump powers \cite{Uesaka2003, Marhic2008}. By employing a resonant optical structure such as a whispering gallery mode (WGM) microring resonator, strong interactions can be achieved with small device footprints, allowing for high conversion efficiencies in a scalable platform \cite{Chang2019, Lu2019, Elshaari2020}. These resonant structures additionally allow for smaller mode areas than bulk crystals, resulting in enhanced nonlinearities at low pump powers \cite{Absil:00, Rodriguez:07}. We consider here a 25 $\mu$m radius ring resonator with coupling of 946 nm and 1550 nm TE modes using separate point waveguide couplers as shown in Fig.\ref{fig:QFC_intro}(b). 

To design a ring resonator geometry for QFC of SiV$^{0}$ emission using FWM-BS, we consider the case of a weak continuous-wave field with frequency $\omega_\text{s}$ as the input signal. When combined with pumps at frequencies $\omega_\text{p1, p2}$ the fields interact to scatter the signal into two new signals, the "idlers", at frequencies  $\omega_\text{i$\pm$} = \omega_\text{p2} \pm |\omega_\text{p1} - \omega_\text{s}|$ where  $\omega_\text{p1}$ is the pump in the 946 nm band, and  $\omega_\text{p2}$ is the pump in the 1550 nm band. In a WGM resonator there are discrete resonances, so it is critical that the converted idlers lie on cavity resonances to achieve appreciable conversion efficiency. Within a given frequency band, the cavity resonances can be described by a Taylor series expansion in orders of the integer mode number $\mu$ relative to some central cavity resonance frequency $\hat{\omega}_\text{0}$ as described by Eq. \eqref{CME_Resonant_Freq_Expansion} \cite{Fujii2020}.

\begin{equation}
    \hat{\omega}_{\mu} = \hat{\omega}_{0} + D_\text{1}\mu + \frac{1}{2}D_\text{2}\mu^{2} + \frac{1}{6}D_\text{3}\mu^{3} + ...
    \label{CME_Resonant_Freq_Expansion}
\end{equation}

\noindent where the coefficients $D_\text{n} \equiv (1/R^{n}) (\partial^{n} \omega / \partial \beta^{n})|_{\omega_0}$ describe the dispersion of the device at the central resonance. The propagation constant is given by $\beta = 2 \pi n_{\text{eff}}(\lambda)/\lambda$, and R is the ring radius. Following the treatment and notation in \cite{li2016efficient}, we denote the cavity resonances nearest the signal, pump, and idler frequencies by $\hat{\omega}_\text{s,p1,p2,i$\pm$}$. For a given pump-signal separation $|\mu|$ in the 946 nm band, the detuning of the converted idlers from their nearest cavity resonance in the 1550 nm band is given by 

\begin{equation}
    \delta \hat{\omega}_{i\pm,|\mu|} = \pm(D_{1}^{1550} - D_{1}^{946})|\mu| + \frac{1}{2} (D_\text{2}^{1550} \mp D_\text{2}^{946}) |\mu|^{2} \pm \frac{1}{6} (D_\text{3}^{1550} - D_\text{3}^{946}) |\mu|^{3} + ...
    \label{Idler_Detuning}
\end{equation}

\noindent The superscripts denote the dispersion coefficients evaluated at the central resonances in the 946 nm and 1550 nm bands respectively. 

To minimize the idler detuning we seek a device geometry with similar dispersion in both bands. Since $D_\text{1} = (1/R)c/n_\text{g}$, where $n_\text{g}$ is the group index, this can be achieved to first order in $|\mu|$ by designing a ring resonator cross-section for which $n_\text{g}$ at 946 nm and 1550 nm is matched. To find such a design, we perform 2D eigenmode simulations of a rectangular GaP waveguide on diamond. The waveguide is defined by its width, $w_\text{y}$, and thickness, $w_\text{z}$, as defined in the previous section. We perform a parameter sweep over the waveguide aspect ratio (AR), defined as $AR = w_\text{y}/w_\text{z}$, and the total cross-sectional area, $A = w_\text{y} w_\text{z}$. In general, we observe that at smaller ARs and smaller areas we achieve greater group index dispersion and better matching between the fundamental TE modes at 946 nm and 1550 nm as shown in Fig.\ref{fig:QFC_intro}(c). The AR and area cannot be reduced arbitrarily, however, because the telecom mode is pushed into cut-off for very small areas and extreme ARs. To avoid cut-off, smaller ARs can be compensated with a larger device area, and smaller device areas can be similarly compensated by a larger AR. As such, there is a tradeoff between AR and area which can be optimized to achieve group index matching without cut-off. In Fig.\ref{fig:QFC_intro}(d) we present simulations for a 25 $\mu$m radius ring resonator, sweeping the resonator width for different thicknesses, from which we select a cross-section geometry with $w_\text{z}$ = 640 nm, $w_\text{y}$ = 500 nm. The group velocity dispersion of the device, shown in Fig.\ref{fig:QFC_intro}(e), exhibits normal dispersion at both 946 nm and 1550 nm.

%QFC Introduction and Ring cross-section selection process

To couple light into the ring resonator, we design two separate point couplers as described in  Fig.\ref{fig:QFC_intro}(f) \cite{Bogaerts2012}. A coupler waveguide with cross-section 640 nm $\times$ 500 nm and 210 nm gap is utilized to couple the mode at 1550 nm with coupling quality factor $Q_\text{C,2}(1550 \text{ nm})=7.71\times10^4$. As the 946 nm mode is much more strongly confined than the telecom mode, there is minimal mode overlap between the 946 nm mode in the coupler and ring resonator, resulting in a severely undercoupled quality factor $Q_\text{C,2}(946\text{ nm})=3.52\times10^8$. To couple the 946 nm signal we utilize a point coupler with cross-section 380 nm $\times$ 300 nm with a 10 nm gap from the resonator, resulting in a coupling quality factor $Q_\text{C,1}(946\text{ nm})=7.64\times10^4$. The thickness of the 946 nm coupler is chosen to match that of the photonic crystal cavity design from the previous section, and a tapering of the width is introduced to achieve the desired coupling Q. Due to the small cross-sectional area, the telecom mode is in cut-off for this coupler \cite{Bi2012}, and as such there is very weak coupling of the telecom mode from the resonator into the waveguide, leading to a coupling quality factor $Q_\text{C,1}(1550\text{ nm}) = 2.70\times10^7$. For both modes we calculate an effective coupling quality factor defined as $1/Q_\text{C} = 1/Q_\text{C,1} + 1/Q_\text{C,2}$ leading to $Q_\text{C}(946\text{ nm})=7.64\times10^4$ and $Q_\text{C}(1550\text{ nm})=7.69\times10^4$.

\begin{figure}[ht!]
\centering\includegraphics[width=\textwidth]{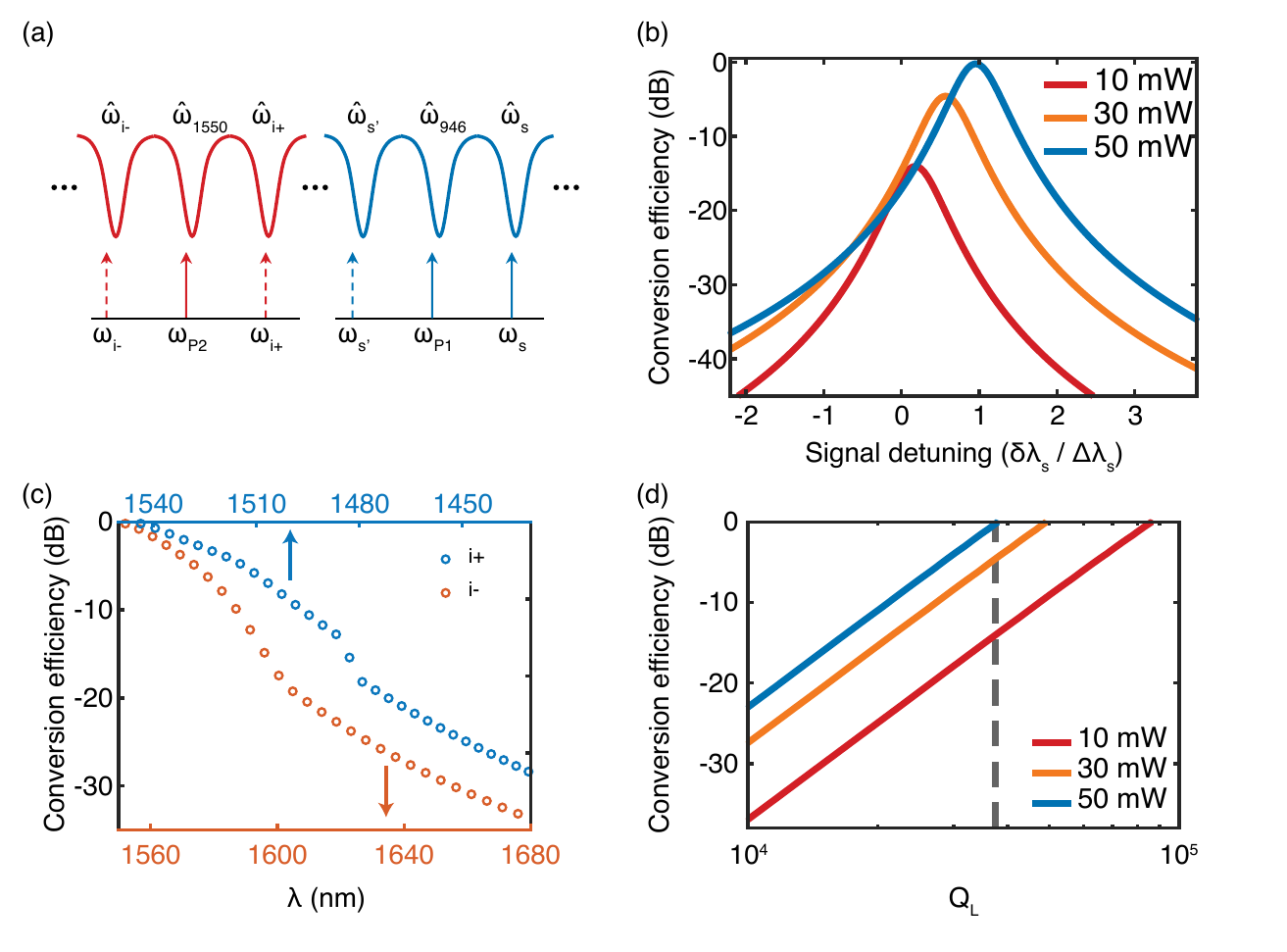}
\caption{(a) Signal,  pump, and idler frequencies are shown as arrows, relative to their nearest cavity resonances. Mixing between the signal, telecom pump, and 946 nm pump leads to conversion into idlers which are offset from the telecom pump by an amount equal to the frequency difference between the signal and 946 nm pump. The frequency shift can occur either to the red (i-) or to the blue (i+). In addition to the idlers, FWM-BS between the signal and 946 nm pump leads to conversion into an auxiliary signal s' with frequency $\omega_\text{s'} = \omega_\text{p1} + (\omega_\text{p1} - \omega_\text{s})$ (b) Conversion efficiency of the i+ idler as a function of signal detuning, $\delta\lambda_\text{s}$, normalized by FWHM of the signal resonance, $\Delta\lambda_\text{s}$, for different total pump power values. The ratio of total power in each pump is kept equal. The signal wavelength is swept near the cavity resonance $\mu=1$. (c) Conversion efficiency bandwidth of the i$\pm$ idlers. As the pump-signal separation $\mu$ is increased, the converted idler frequencies are correspondingly shifted further from the 1550 nm pump. The top (bottom) x-axis indicates the conversion efficiency of the i- (i+) idler at a given output wavelength. Conversion efficiency drops for larger $\mu$ as the idlers become increasingly detuned from their nearest cavity resonance. The conversion efficiency for i$\pm$ is non-symmetric because the resonance spacing is not identical at wavelengths greater than 1550 nm and wavelengths smaller than 1550 nm. (d) Conversion efficiency for the i+ idler as a function of the ring total quality factor $Q_\text{L}$ and total pump power, under the assumption of critical coupling at both 946 nm and 1550 nm for a pump-signal separation $\mu=1$. The vertical dashed line indicates a loaded quality factor $Q_\text{L}=3.75 \times 10^{4}$.}
\label{fig:QFC_CME_Results}
\end{figure}

To evaluate the FWM-BS conversion efficiency in our designed coupler-resonator system, we study analytic coupled mode equations \eqref{CME_Signal} - \eqref{CME_i_minus} for the fields inside a resonator with a $\chi^{(3)}$ as described in detail in \cite{li2016efficient}. In addition to our pumps, signal, and idlers, we consider mixing between the signal and the 946 nm band pump which scatters the signal into an auxiliary signal $\omega_\text{s'} = 2\omega_\text{p1} - \omega_\text{s}$ as shown in Fig.\ref{fig:QFC_CME_Results}(a). 

%Include input signal power here with coupling coefficient then describe in text
\begin{equation}
    t_\text{R} \frac{dE_\text{s}}{dt} = - (\alpha_\text{p1} + i \Delta \phi_\text{s})E_\text{s}  + i \gamma_\text{p1}LE^{2}_\text{p1}E^{*}_\text{s'} + i2\gamma_\text{p1}LE_\text{p1}(E_\text{p2}E^{*}_\text{i-} + E_\text{p2}^{*}E_\text{i+}) + i \sqrt{\theta_\text{1}P_\text{s}}
    \label{CME_Signal}
\end{equation}

\begin{equation}
    t_\text{R} \frac{dE_\text{s'}}{dt} = - (\alpha_\text{p1} + i \Delta \phi_\text{s'})E_\text{s'}  + i \gamma_\text{p1}LE^{2}_\text{p1}E^{*}_\text{s} + i2\gamma_\text{p1}LE_\text{p1}(E_\text{p2}E^{*}_\text{i+} + E_\text{p2}^{*}E_\text{i-})
    \label{CME_Aux}
\end{equation}

\begin{equation}
    t_\text{R} \frac{dE_\text{i+}}{dt} = - (\alpha_\text{p2} + i \Delta \phi_\text{i+})E_\text{i+}  + i \gamma_\text{p2}LE^{2}_\text{p2}E^{*}_\text{i-} + i2\gamma_\text{p2}LE_\text{p2}(E_\text{p1}E^{*}_\text{s'} + E_\text{p1}^{*}E_\text{s})
    \label{CME_i_plus}
\end{equation}

\begin{equation}
    t_\text{R} \frac{dE_\text{i-}}{dt} = - (\alpha_\text{p2} + i \Delta \phi_\text{i-})E_\text{i-}  + i \gamma_\text{p2}LE^{2}_\text{p2}E^{*}_\text{i+} + i2\gamma_\text{p2}LE_\text{p2}(E_\text{p1}E^{*}_\text{s} + E_\text{p1}^{*}E_\text{s'})
    \label{CME_i_minus}
\end{equation}

\noindent Where $\text{p}1$ and $\text{p}2$ denote the pumps at 946 nm and 1550 nm respectively, $L=2 \pi R$ is the cavity round-trip length, $t_\text{R}=2 \pi / D_\text{1}$ is the round-trip time, and $\alpha_m=\hat{\omega}_\text{m} t_\text{R} / 2Q_\text{L,m}$ is the cavity loss for a resonance $m$ with loaded quality factor $Q_\text{L,m}$. The effective nonlinear parameter is given by $\gamma=n_2\omega/cA_\text{eff}$. $P_\text{s}$ is the power of the signal at the waveguide input which is coupled into the resonator with power coupling ratio $\theta_\text{1}=\hat{\omega}_\text{p1}t_\text{R}/Q_\text{C}(946 \text{ nm})$. Here we neglect nonlinear effects arising from the signal and auxiliary signal. The round trip loss, nonlinear parameters, and power coupling ratios are assumed to be constant within the 946 nm and 1550 nm bands. The $\Delta \phi_\text{m}$ terms describe the effective detuning of the modes from their nearest cavity resonance, including the effects of Kerr nonlinear frequency shifts of the cavity resonances caused by the strong pumps, given by:

\begin{equation}
    \Delta \phi_\text{s,s'} = (\hat{\omega}_\text{s,s'} - \omega_\text{s,s'})t_\text{R} - 2\gamma_\text{p1}L(|E_\text{p1}|^{2} + |E_\text{p2}|^{2})
    \label{CME_Detuning_signals}
\end{equation}

\begin{equation}
    \Delta \phi_\text{i$\pm$} = (\hat{\omega}_\text{i$\pm$} - \omega_\text{i$\pm$})t_\text{R} - 2\gamma_\text{p2}L(|E_\text{p1}|^{2} + |E_\text{p2}|^{2})
    \label{CME_Detuning_idlers}
\end{equation}

To calculate $\gamma_\text{p1,p2}$ we first perform 2D eigenmode simulations of the ring cross-section to extract the TE field profiles at 946 nm and 1550 nm. We then calculate the effective mode area, A$_\text{eff}$, as defined in Eq. \eqref{CME_Aeff} where the integral in the denominator is evaluated over the resonator cross-section only: 

\begin{equation}
    A_\text{eff} = \frac{\left( \int \int \epsilon_\text{r}(x,y) |E(x,y)|^{2} dxdy \right)^{2}} 
                    {\int \int_\text{Ring} \epsilon_\text{r}^{2}(x,y) |E(x,y)|^{4} dxdy}
    \label{CME_Aeff}
\end{equation}

We determine the central resonance in the 946 nm and 1550 nm bands via 2D eigenmode simulations of the effective index of the TE modes, yielding resonances at $\hat{\omega}_\text{p1} = 2 \pi c / 946.6 \text{ nm}$ and $\hat{\omega}_\text{p2} = 2 \pi c /1547.8 \text{ nm}$ respectively. We solve  Eq. \eqref{CME_Signal} - \eqref{CME_i_minus} in the steady state and compute the conversion efficiency, $\eta_\text{\text{conv}}$, in terms of idler photon flux at the output of the coupling waveguide relative to the signal photon flux at the input:

\begin{equation}
    \eta_\text{conv} = \frac{\theta_\text{2} |E_\text{i$\pm$}|^{2}/\hbar \omega_\text{i$\pm$}}{P_\text{s}/ \hbar \omega_\text{s}}
    \label{CME_Conversion_Efficiency}
\end{equation}

\noindent Where $\theta_\text{2}=\hat{\omega}_\text{p2}t_\text{R}/Q_\text{C}(1550 \text{ nm})$. Table \ref{table:QFC_constants} summarizes the constants used in the analysis.

\begin{table}[ht!]
\centering
\resizebox{\textwidth}{!} {\begin{tabular}{ c c c c c c c c }
    \hline
    Wavelength Band & $D_1/2\pi$ (MHz) & $D_2/2\pi$ (MHz) & $D_3/2\pi$ (MHz) & $Q_i$ ($10^4$) & $Q_\text{C}$ ($10^4$) & $n_2$ ($10^{-17} m^2 W^{-1}$) & $A_\text{eff}$ ($\mu m^2$)\\
    \hline
    946 nm & 531 & -122 & -0.964 & 7.5 & 7.64 & 0.85 \cite{Grinblat2019} & 0.196 \\  
    1550 nm & 531 & -44.2 & 13.1 & 7.5 & 7.69 & 1.1 \cite{Wilson2020} & 0.292 \\
    \hline
\end{tabular}}
\caption{Constants used for solving the coupled mode equations}
\label{table:QFC_constants}
\end{table}

%Results of coupled-mode equation simulations of frequency conversion process

In Fig.\ref{fig:QFC_CME_Results}(b) we analyze the conversion efficiency for the i+ idler as a function of normalized signal detuning from the $\mu=1$ cavity resonance for different total pump powers.  We assume a ring with intrinsic quality factor $Q_\text{i}=7.5 \times 10^{4}$, yielding loaded quality factors $Q_\text{L}^{946}=3.78 \times 10^{4}$ and $Q_\text{L}^{1550}=3.80 \times 10^{4}$ using the coupler design from Fig.\ref{fig:QFC_intro}(f). As GaP microring resonators with loaded quality factors in excess of $10^{5}$ have been experimentally demonstrated in the telecom band \cite{Wilson2018, Wilson2020} we believe this to be a conservative estimate. We achieve a near-unity conversion efficiency of $-0.21$ dB for i+ at a total pump power of $50$ mW and signal detuning of $0.95\Delta\lambda_s$, where $\Delta\lambda_{s}$ is the full width at half maximum (FWHM) of the signal resonance. At higher pump powers the signal cavity resonance experiences a nonlinear Kerr shift which requires the signal frequency to be correspondingly shifted to achieve maximal conversion efficiency.

Fig.\ref{fig:QFC_CME_Results}(c) shows the conversion efficiency into both idlers as a function of signal-pump separation $\mu$. At larger values of $\mu$ conversion efficiency is reduced due to dispersion in the resonator: the resonances are not equally spaced with the same magnitude for both wavelength bands, and this difference increases with larger detuning from the central pump resonance. The efficiency for i$\pm$ is not symmetric because the dispersion differs at wavelengths below the central resonance as compared to wavelengths above the central resonance. As such, the detuning for i- is not the same as the detuning for i+.

In Fig.\ref{fig:QFC_CME_Results}(d) we analyze conversion efficiency into i+ as a function of the ring loaded quality factor $Q_\text{L}$ for different total pump powers under the assumption of critical coupling $Q_\text{C}=Q_\text{i}$. For a total pump power of 30 mW, we predict unity conversion efficiency for $Q_\text{L} = 4.9 \times 10^4$ at the optimal signal detuning, and for 10 mW we require $Q_\text{L} = 8.6 \times 10^4$. Moving to lower pump powers is favorable for reducing the effects of nonlinear and thermal frequency shifts of the cavity resonances which can make it difficult for the pumps and signal to be simultaneously on-resonance with the cavity modes \cite{li2016efficient}. In addition to the assumptions of critical coupling, we neglect pump depletion and parasitic nonlinear processes such as higher order idler generations in the 1550 nm band. These processes serve to deplete the primary idlers and thereby reduce the conversion efficiency \cite{li2016efficient}.

\section{Discussion} \label{section:discussion}
%Discuss why GaAs QFC doesnt work for SiV0 here

In this paper, we have demonstrated the potential of hybrid III-V diamond platforms for quantum networks based on SiV$^0$ centers in diamond. We show that a heterogeneously integrated nanobeam cavity can enhance the photon emission from a single SiV$^0$ center, and the emitted photon can be subsequently converted to the telecommunication C-band on chip via a FWM-BS scheme, providing an efficient interface between the SiV$^0$ center and photons at telecommunication wavelengths. In a fully integrated platform, the system will require joint optimization between Purcell enhancement and frequency conversion. Specifically, there will be a trade-off between the photon collection efficiency into the cavity (i.e. $P/(P+1)$) and the maximum conversion efficiency achievable with the frequency converter because the optical linewidth will radiatively broaden beyond the bandwidth of the microresonator-based frequency converter \cite{Singh2018}. For the system presented here, the linewidth of SiV$^0$ emission (88 MHz) is narrow compared to the loaded linewidth of the ring resonator in Section \ref{section:QFC} ($\sim 8.4$ GHz), which allows for a Purcell factor of around 100 before the broadened optical linewidth reaches the bandwidth of the ring resonator. An example of such a device would utilize a large number of cavity holes (greater than 20) to suppress far-field scattering and a small number of mirror segments (less than 5) to allow for decay into the waveguide mode, with a total Purcell factor of less than 100 and a collection efficiency over 90$\%$.

The proposed platform can be experimentally realized using widely available existing fabrication techniques. Here, we comment on practical implementation of the hybrid platform. Heterogeneous integration of quantum photonic elements with different functionalities has been demonstrated in a wide variety of material platforms \cite{Elshaari2020}. Among different fabrication approaches to assemble the elements, epitaxial lift-off, in which the III-V membrane layer is released from the heteroepitaxially-grown substrate through selective wet etching and subsequently transferred onto diamond \cite{gould2016efficient}, and transfer printing, where suspended photonic devices can be picked up and transferred using a rubber stamp \cite{katsumi2018transfer, dibos2018atomic}, are promising routes towards large-scale integration. The two techniques differ in the method for aligning photonic elements to the color centers. With the epitaxial lift-off approach, the nanobeam cavity and the micro-resonator for QFC can be defined by lithography with sub-50 nm alignment accuracy. In principle, this would enable large-scale photonic circuit elements with more complex functionality \cite{gould2016large}. However, the applicability of this technique may be limited by the inhomogeneous distribution of SiV$^0$ centers and the spectral variation of nanobeam cavities resulting from fabrication imperfections. On the other hand, transfer printing allows the integration of pre-selected color centers in diamond with pre-characterized photonic devices. Previous work has demonstrated similar alignment accuracy to the epitaxial lift-off approach \cite{katsumi2018transfer}. Despite the challenges in achieving large-scale photonic device integration, this method promises a robust and reproducible way to couple individual SiV$^0$ centers to high-performance photonic devices.

While our hybrid platform design targets SiV$^0$ centers in diamond with an emission wavelength of 946 nm, our general design approach for designing the bandgap for the nanobeam cavity and the dispersion for miroresonator-based FWM-BS can be applied to a wide variety of color centers in diamond and other host materials. For example, the GaP-on-diamond platform could be applied to the negatively charged silicon vacancy centers in diamond due to the wide transparancy window of GaP. Furthermore, while the proposed FWM-BS QFC scheme is not feasible for SiV$^0$ in the GaAs-on-diamond platform due to the band edge absorption feature which dominates the index dispersion at 946 nm, GaAs could be used for color centers in silicon carbide that emit in the near-infrared wavelength ranges \cite{diler2020coherent, wolfowicz2020vanadium}.

\section*{Acknowledgements}
The authors would like to thank Kartik Srinivasan, Alejandro Rodriguez, Zi-Huai Zhang, and Paul Stevenson for helpful discussions about the photonic crystal cavity design and FWM-BS scheme. This work was primarily supported by DARPA under Young Faculty Award (award number D18AP00047), and work on the neutral silicon vacancy center was supported by the NSF under the EFRI ACQUIRE program (grant 1640959) and the Air Force Office of Scientific Research under award number FA9550-17-0158. M. R. and J. D. T. were additionally supported by the Air Force Office of Scientific Research (grant FA9550-18-1-0081). D. H. was supported by a National Science Scholarship from A*STAR, Singapore. A. A. was supported by a Post Graduate Scholarship from the Natural Sciences and Engineering Research Council of Canada.

\section*{Disclosures}
The authors declare no conflicts of interests.

\bibliography{reference}

\end{document}